\documentstyle[preprint,aps]{revtex}

\begin{document}
\title{An Investigation of MBBA nematic within Generalized Maier-Saupe Theory}
\author{O. Kayacan\thanks{
e-mail: ozhan.kayacan@bayar.edu.tr}}
\address{Department of Physics, Faculty of Art and Science, Celal Bayar University,\\
Muradiye/Manisa-TURKEY}
\maketitle

\begin{abstract}
\qquad Generalized Maier-Saupe theory (GMST) within Tsallis thermostatistics
(TT) has been used to investigate the second rank order parameter in a
nematic liquid crystal, N-p-methoxybenzylidene p-n-butylaniline (MBBA).
Also, the free energy has been investigated at the nematic-isotropic phase
transition temperature and the effect of the nonextensivity has been
demonstrated. \newline

\noindent {PACS Number(s):}~ 05.20.-y, 05.70.-a, 61.30.Cz, 61.30. Gd.\newline

\end{abstract}

\newpage

\section{Introduction}

\qquad So far, many statistical theories of nematic liquid crystals have
been proposed. One of the most popular theories of the nematics is
Maier-Saupe (MST) [1,2] which is a mean field study. Although MST is
successful to describe the nematic-isotropic transition for some nematics,
it is not sufficent for some others. In this study, we used generalized
Maier-Saupe theory (GMST) to study a nematic liquid crystal,
N-p-methoxybenzylidene p-n-butylaniline($MBBA$), where MST is known to fail. Most molecular theories of nematic liquid crystals assume that the constituent molecules are cylindrically symmetric and MST is one of these molecular theories. This may be a useful approximation. However, the molecules of real nematogens are lower symmetry. Therefore the dependence of the orientational properties of the uniaxial mesophase on the deviation from molecular cylindrical symmetry was calculated from the series expansion of the pseudopotential [3]. This points out that the entropy of the system according to MST is of course not appropriate for the nematics. Because the molecules have lower symmetry, we should use another entropy definition, i.e. another thermostatistics. In this context, I use GMST within TT to investigate the nematics. The aim of this study is to show that the TT may be an appropriate one of these thermostatistics. 

\qquad We start describing the Tsallis thermostatistics (TT). Since the
paper by Tsallis [4], there has been a growing tendency to nonextensive
statistical formalism. It has been shown that this formalism is useful,
because it provides a suitable theoretical tool to explain some of the
experimental situations, where standard thermostatistics seems to fail, due
to the presence of long-range interactions, or long-range memory effects, or
multi-fractal space-time constraints. The axioms of TT are the following:

TT has been applied to various concepts of thermostatistics and achieved in
solving some physical systems, where Boltzmann-Gibbs statistics is known to
fail [5]. Recently, MST has been generalized within TT and GMST has been
applied to p-azoxyanisole ($PAA$) in [6] in which Kayacan et al. used the
second choice for the internal energy constraint which will be given below.

TT considers three possible choices for the form of a nonextensive
expectation value. These choices have been studied in [7] and applied to two
systems; the classical harmonic oscillator and the quantum harmonic
oscillator. In that study, Tsallis et al. studied three different
alternatives for the internal energy constraint. The first choice is the
conventional one and used in [3] by Tsallis

\begin{equation}
\sum_{i=1}^{W}p_{i}\varepsilon _{i}=U_{q}^{(1)}.
\end{equation}
The second choice is given by

\begin{equation}
\sum_{i=1}^{W}p_{i}^{q}\varepsilon _{i}=U_{q}^{(2)}
\end{equation}
and regarded as the canonical one. Both of these choices have been applied
to many different systems in the last years [8]. However both of them have
undesirable difficulties. The third choice for the internal energy
constraint is

\begin{equation}
\frac{\sum_{i=1}^{W}p_{i}^{q}\varepsilon _{i}}{\sum_{i=1}^{W}p_{i}^{q}}%
=U_{q}{}^{(3)}.
\end{equation}
This choice is commonly considered to study physical systems because it is
the most appropriate one, and is denoted as the Tsallis-Mendes-Plastino
(TMP) choice. $q$ index is called the entropic index and comes from the
entropy definition [3],

\begin{equation}
S_{q}=k\frac{1-\sum_{i=1}^{W}p_{i}^{q}}{q-1}
\end{equation}

where $k$ is a constant, $\sum_{i}p_{i}=1$ is the probability of the system
in the $i$ microstate, $W$ is the total number of configurations. In the
limit $q\rightarrow 1$, the entropy reduces to the well-known
Boltzmann-Gibbs (Shannon) entropy.

\qquad The optimization of $S_{q}$ leads to

\begin{equation}
p_{i}^{(3)}=\frac{\left[ 1-(1-q)\beta (\varepsilon
_{i}-U_{q}^{(3)})/\sum_{j=1}^{W}(p_{j}^{(3)})^{q}\right] ^{\frac{1}{1-q}}}{%
Z_{q}^{(3)}}
\end{equation}

with

\begin{equation}
Z_{q}^{(3)}=\sum_{i=1}^{W}\left[ 1-(1-q)\beta (\varepsilon
_{i}-U_{q}^{(3)})/\sum_{j=1}^{W}(p_{j}^{(3)})^{q}\right] ^{\frac{1}{1-q}}.
\end{equation}
This equation is an implicit one for the probabilities $p_{i}$. Therefore
the $normalized$ $q-expectation$ $value$ of an observable is defined as

\begin{equation}
A_{q}=\frac{\sum_{i=1}^{W}p_{i}^{q}A_{i}}{\sum_{i=1}^{W}p_{i}^{q}}%
=\left\langle A_{i}\right\rangle _{q}
\end{equation}

where $A$ denotes any observable quantity which commutes with the
Hamiltonian. This expectation value recovers the conventional expectation
one when $q=1$. As mentioned above, Eq.(7) is an implicit one and in order
to solve this equation, Tsallis et al. suggest two different approaches; ''$%
iterative$ $procedure$'' and ''$\beta \rightarrow \beta ^{^{\prime }}$''
transformation.

This study is the second one which use the GMST to study the nematics.
Therefore another aim of this study is to take a step towards the enlargement of
the use of TT in liquid crystals and application domain of GMST by using the
third choice for the internal energy constraint. It is worthwhile to note
that in the previous paper [6], Kayacan et al. used the second choice
whereas the third choice is considered in the present study. It is
well-known from the literature that the first two choices have some
disadvantages, so we use the third choice which is the most appropriate one,
as mentioned above.

Also, there are some extensions of the MST, including molecular field terms
into the potential energy, using BG statistics [9,10]. In this study,
long-range interactions have been taken into account for $q\neq 1$, without
including the terms into the interaction potential energy, and the
experimental data have been successfully explained for second rank order
parameter of $MBBA$, within TT, with a small departure from the standard
theory.

\section{Results and Discussion}

\qquad MST assumes the intermolecular potential energy as

\begin{equation}
u_{i}=-\frac{A}{V^{2}}P_{2}\left( \frac{3\cos ^{2}\theta _{i}-1}{2}\right)
\end{equation}
which bases on dispersion forces. In this equation, $A$ is a constant
independent of pressure, volume and temperature, $V$ is molar volume, and $%
\theta _{i}$ is the angle between the long molecular axis and the preferred
axis, $P_{2}$ is the second rank order parameter and given by

\begin{equation}
P_{2}=\left\langle \frac{3\cos ^{2}\theta _{i}-1}{2}\right\rangle =s
\end{equation}

in which $P_{2}$ denotes the Legendre polynomial with $l=2$. The generalized
second rank order parameter is calculated from

\begin{equation}
\left\langle P_{2}\right\rangle _{q}=\frac{\int_{0}^{1}P_{2}(x_{i})%
\,p_{i}^{q}d(x_{i})}{\int_{0}^{1}\,p_{i}^{q}d(x_{i})}
\end{equation}

within TT, where $x_{i}$ denotes the direction cosine of the long axis of an
individual molecule with respect to the nematic axis.

\qquad Now let us treat the Helmholtz free energy of the system and the
equilibrium condition within TT. The Helmholtz free energy is written as

\begin{equation}
F_{q}=U_{q}-TS_{q}
\end{equation}

where

\begin{equation}
U_{q}=\frac{N}{2}\frac{\int_{0}^{1}u_{i}(x)\,\,\,p_{i}^{q}\,dx_{i}}{%
\int_{0}^{1}p_{i}^{q}\,dx_{i}}.
\end{equation}
Thus the entropy of the system can be expressed as

\begin{equation}
S_{q}=k\beta U_{q}+k\log _{q}(Z_{q}^{(3)})
\end{equation}

in which $Z_{q}^{(3)}$ is given by Eq.(6). If substituting Eqs.(12) and (13)
into Eq.(11), the Helmholtz free energy associated with the system is
obtained. The condition for the equilibrium in the nematic-isotropic liquid
phase transition reads

\begin{equation}
\left( \frac{\partial F_{q}}{\partial s_{q}}\right) _{V,T}=0
\end{equation}

and this condition may be called "the consistency relation".

\qquad G. Sigaud et al. reported the measurements of the magnetic anisotropy
and $\left\langle P_{2}\right\rangle $ data of $MBBA$ [11]. Fig.(1) shows
the second rank order parameter as a function of temperature. The curves are plotted in nematic range of $MBBA$. It is seen
that the fit to $\left\langle P_{2}\right\rangle $ data is very good for $%
q=0.978$, whereas the result of MST is very poor. The variation of the free
energy as a function of the order parameter at nematic-isotropic phase
transition temperature is shown in Fig.(2). It is seen that the free energy
has two minima, one at $s=0$ representing the isotropic phase, and the other
one at $s=s_{c}$ corresponding to the nematic phase. The two states have
equal free energies and at this temperature, a discontinuous transition
takes place. The minima corresponding to the nematic phase are at $%
s_{c}=0.43 $ for $q=1$ (MST) and at $s_{c}=0.35$ for $q=0.978$. This result
is in far better agreement with the experimental one, $s_{c}=0.32$ [12] than
MST.

\qquad It is worthwhile to imply that the experimental values of $%
\left\langle P_{2}\right\rangle $ are higher than those predicted by MST for
most nematics[13]. The most-known exceptions are $PAA$ and $MBBA$ for which
MST predictes higher values of $\left\langle P_{2}\right\rangle $ than the
experimental ones. Therefore, it seems that the $\left\langle
P_{2}\right\rangle $ curves of most nematics might be expected to be fitted
appropriately with $q$ values greater than unity. Since the nematics are a
class of the liquid crystals, this case is the expected one.

\section{Conclusion}

\qquad It is seen from this study that MST can be improved by using TT. In
this manner, GMST explains the variation of the second rank order parameter
and the value of the order parameter at the transition temperature for $MBBA$%
. It is seen from the present and previous studies on some kinds of nematics
that GMST is a useful tool to study the nematics. Some disadvantages of MST
can be improved by using TT.

\section*{Acknowledgements}

The author would like to thank U. Tirnakli for fruitful comments and
suggestions.\newline

\section*{References}

[1] W. Maier, A. Saupe, Z. Naturforsch. 13a, 564 (1959).

[2] W. Maier, A. Saupe, Z. Naturforsch. 15a, 287 (1960).

[3] G.R. Luckhurst and C. Zannoni, Mol. Liq. 30, 1345 (1975).

[4] C. Tsallis, J. Stat. Phys. 52, 479 (1988).

[5] S. Abe, Y. Okamoto, Nonextensive statistical mechanics and its
applications, Series Lectures Notes in Physics, Berlin: Springer-Verlag,
2001.

[6] O. Kayacan, F. B\"{u}y\"{u}kk\i l\i \c{c}, D. Demirhan, Physica A 301,
255 (2001).

[7] C. Tsallis, R.S. Mendes, A.R. Plastino, Physica A 261, 534 (1998).

[8] see http://tsallis.cat.cbpf.br for an updated bibliography.

[9] R.L. Humphries, P.G. James, G.R. Luckhurst, J. Chem. Soc., Faraday
Trans. 2 68, 1031 (1972).

[10] S. Chandrasekhar, N.V. Madhusudana, Mol. Cryst. Liq. Cryst. 10, 151
(1970).

[11] G. Sigaud, H. Gasparoux, J. Chim. Phys. Physicochim. Biol. 70, 699
(1973).

[12] A. Pines, J.J. Chang, Phys. Rev. A 10, 946 (1974).

[13] A. Saupe, Angew. Chem. Internat. Edit. 7, 97 (1968).

\newpage

{\bf Figure Caption} \newline

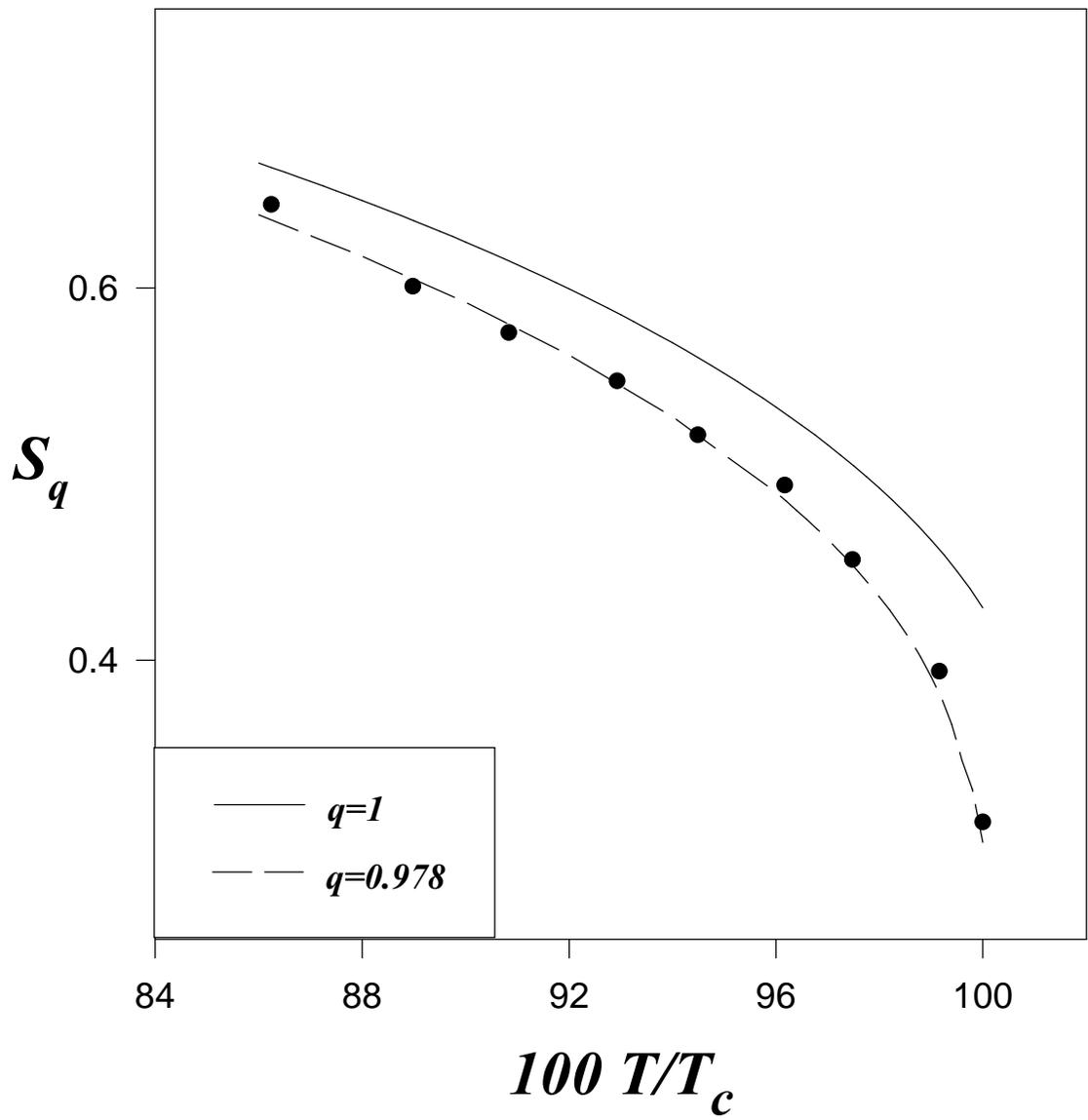
Figure 1. Orientational order parameter as a function of temperature in MBBA
for $q=0.978$ and $q=1$. The curves are plotted in nematic range of $MBBA$. Filled circles represent the experimental data[11].%
\newline

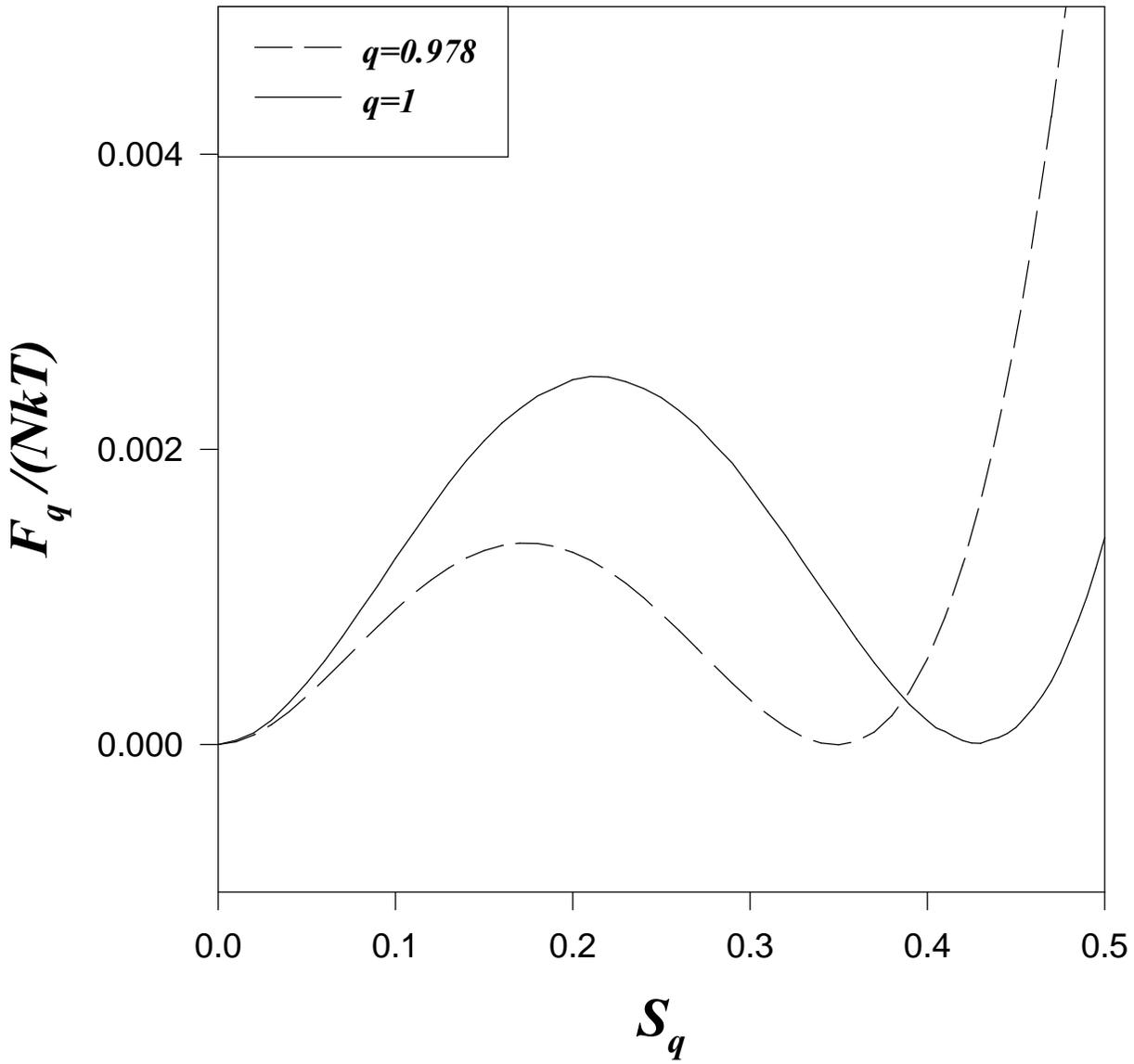
Figure 2. The Helmholtz free energy as a function of generalized order
parameter in MBBA for $q=0.978$ and $q=1$.\newline

\end{document}